\documentclass[conference, 10pt]{IEEEtran}

\usepackage{cite}
\usepackage{amsmath,amssymb,amsfonts}
\usepackage{algorithmic}
\usepackage{graphicx}
\usepackage{textcomp}
\usepackage{xcolor}
\usepackage{breqn}
\usepackage{algorithmic}
\usepackage{mathtools}
\usepackage{comment}
\usepackage[flushleft]{threeparttable}
\usepackage{booktabs, array, dcolumn}

\ifCLASSOPTIONcompsoc
\usepackage[caption=false,font=normalsize,labelfon
t=sf,textfont=sf]{subfig}
\else
\usepackage[caption=false,font=footnotesize]{subfi
g}
\fi

\def\BibTeX{{\rm B\kern-.05em{\sc i\kern-.025em b}\kern-.08em
    T\kern-.1667em\lower.7ex\hbox{E}\kern-.125emX}}
\begin{document}

\title{An End-to-End Authentication Mechanism for Wireless Body Area Networks
}

\author{\IEEEauthorblockN{Mosarrat Jahan, Fatema Tuz Zohra, Md. Kamal Parvez, Upama Kabir, Abdul Mohaimen Al Radi, Shaily Kabir}
\IEEEauthorblockA{Department of Computer Science and Engineering, University of Dhaka, Dhaka, Bangladesh\\
Email: mosarratjahan@cse.du.ac.bd, fatema.zohra.2214@gmail.com, kamalparvez02@gmail.com, upama@cse.du.ac.bd,\\ 2018-925-300@student.cse.du.ac.bd, shailykabir@cse.du.ac.bd}}

\maketitle

\begin{abstract}
Wireless Body Area Network (WBAN) ensures high-quality healthcare services by endowing distant and continual monitoring of patients' health conditions. The security and privacy of the sensitive health-related data transmitted through the WBAN should be preserved to maximize its benefits. In this regard, user authentication is one of the primary mechanisms to protect health data that verifies the identities of entities involved in the communication process. Since WBAN carries crucial health data, every entity engaged in the data transfer process must be authenticated. In literature, an end-to-end user authentication mechanism covering each communicating party is absent.  Besides, most of the existing user authentication mechanisms are designed assuming that the patient's mobile phone is trusted. In reality, a patient's mobile phone can be stolen or comprised by malware and thus behaves maliciously. Our work addresses these drawbacks and proposes an end-to-end user authentication and session key agreement scheme between sensor nodes and medical experts in a scenario where the patient's mobile phone is semi-trusted.  We present a formal security analysis using BAN logic. Besides, we also provide an informal security analysis of the proposed scheme. Both studies indicate that our method is robust against well-known security attacks. In addition, our scheme achieves comparable computation and communication costs concerning the related existing works. The simulation shows that our method preserves satisfactory network performance.
\end{abstract}


\section{Introduction} \label{introduction}
Wireless Body Area Network (WBAN) promotes healthcare services by enabling continuous remote monitoring of the patients. To do so, it forms a short-range wireless network using the sensor nodes associated with the human body, responsible for monitoring and collecting different physiological data and communicating those data to healthcare services through the wireless signal. Hence, WBAN eliminates the need of the patients to frequently visit hospitals and turns the laborious task of healthcare givers more systematic. Especially, WBAN is beneficial for monitoring elderly patients and patients suffering from chronic conditions.  
	
Nevertheless, wide deployment of WBAN is subject to concern due to various security and privacy issues caused mainly by the involvement of resource-constrained sensor nodes \cite{wang2015preserving, saeed2018remote}. Moreover, WBAN transfers highly sensitive health-related data \cite{baker2017internet, al2018context}. Therefore, the development of lightweight and rigorous security mechanisms is essential for the practical realization of WBAN. In this regard, user authentication is a predominant mechanism to confirm the identities of participating nodes and combat unauthorized access to patients' data. 

Although current research works address the user authentication mechanism of WBAN \cite{saeed2018remote, jegadeesan2020epaw, li2018secure}, they do not take into consideration various communication among the WBAN entities. Usually, WBAN follows a centralized two-hop WBAN architecture \cite{li2018secure, li2017anonymous, kompara2019robust}. Here, sensor nodes collect physiological data such as blood glucose level, pulse rate, body temperature, and heart rate \cite{kompara2019robust} and transmit to an intermediate node, generally the mobile phone associated with a patient. This communication is known as \textit{intra-BAN} communication \cite{li2018secure}. In addition, the intermediate node transfers data to a hub node, and this communication is known as \textit{inter-BAN} communication \cite{li2018secure}. Finally, the hub node transfers data to the health service providers using \textit{beyond-BAN} communication \cite{li2018secure}. In literature, most works \cite{li2018secure, konan2019secure, chen2019analysis} proposed authentication mechanisms for the inter-BAN communication without providing any clue regarding the secure communication mechanism between the sensor nodes and the patient's mobile phone. Only \cite{wazid2017novel, abiramy2019secure} mentioned a key establishment mechanism for the intra-BAN part while proposing an authentication mechanism for the inter-BAN communication. As sensitive health data passes through each WBAN entity, an end-to-end authentication covering each communication between the WBAN entities is essential. Although \cite{li2017anonymous, kompara2019robust} proposed an authentication mechanism between the sensor nodes and hub node, these schemes can optionally utilize the patient's mobile phone as a forwarder node, and the authors considered the mobile phone to be completely trusted. In reality, sensor nodes in WBAN use an intermediate resource-rich device such as a patient's smartphone and smartwatch to reduce energy overhead to transmit to a distant entity \cite{li2017anonymous, wazid2017novel}. Therefore, in a realistic scenario patient's associated mobile device should also participate in the authentication process. Moreover, a patient's mobile phone can be stolen or affected by malware that secretly eavesdrops on valuable information. Therefore, the assumption of a completely trusted mobile phone is not practical. 

To address these shortcomings, we extend Al-Turjman and Alturjman's scheme \cite{al2018context} by incorporating the patient's mobile phone in the authentication process and considering the mobile phone as a semi-trusted entity. In particular, the following contributions are made in this paper:

\begin{itemize}
    \item We present an end-to-end user authentication and session key establishment mechanism to support secure communication between the sensor nodes connected to patients' bodies and health experts. This scheme covers intra-BAN, inter-BAN, and beyond-BAN transmission in a setting where the patient's mobile phone is semi-trusted.
    
    \item We present a rigorous security analysis of the proposed scheme using widely accepted BAN logic. Besides, we also give an informal security analysis of the proposed scheme. 
    	
    \item We demonstrate the performance of the proposed scheme concerning the other related works using computation and communication costs.
	
	\item We implement the proposed scheme using NS-3 \cite{riley2010ns} simulator and assess the effect of the proposed scheme on various network parameters.
\end{itemize}

The remaining paper is organized as follows. Section \ref{related_work} summarizes the related works on the WBAN authentication mechanism. Besides, Section \ref{system_model} presents the system model of our proposed scheme, while Section \ref{proposed_scheme} provides a comprehensive description of the proposed scheme. Section \ref{security_analysis} discusses the security features of the proposed scheme. In addition, Section \ref{protocol_analysis} offers formal security proof using BAN logic, and section \ref{comparative_study} presents a comparative performance analysis of the proposed scheme. Section \ref{practical_impact} illustrates the effect of the proposed scheme on network performance. Lastly, Section \ref{conclusion} concludes the paper.
\section{Literature Review}\label{related_work}
Baker et al. \cite{baker2017internet} presented a comprehensive study on the application of the Internet of Things (IoT) in the healthcare system and highlighted the recent research works in this direction. This study identifies the lack of research on providing treatment in emergencies. Also, it indicates the insufficiency of research on security schemes that covers end-to-end IoT-based healthcare systems. For example, Saeed et al. \cite{saeed2018remote} presented a lightweight and anonymous user authentication scheme between a WBAN sensor and the application provider using an online/offline certificate-less signature mechanism. Hence, this scheme does not authenticate every entity of WBAN. Besides, Abiramy and Sudha \cite{abiramy2019secure} proposed a lightweight inter-BAN authentication scheme between the patient's mobile device and application providers. Further, this scheme creates a group key to support secure data transfer operation among the mobile terminal and sensor nodes. Hence, every sensor node and the mobile terminal can listen to messages interchanged by other sensors. Wazid et al. \cite{wazid2017novel} handled this shortcoming by establishing pairwise secret keys between the implanted sensors and the patients' mobile phone. Further, the authors proposed a three-factor remote user authentication mechanism between a doctor and a patient's mobile phone. Similarly, Li et al. \cite{li2018secure} proposed an authentication mechanism between a patient's mobile phone and the medical expert in a three-phase mobile healthcare system. Besides, Konan and Wang \cite{konan2019secure} introduced an efficient authentication scheme between the smartphone of a patient and the application provider. Moreover, the authors proposed a batch authentication process to reduce the computation and communication costs. On the other hand, Arfaoui et al. \cite{arfaoui2019context} proposed a context-aware anonymous intra-BAN authentication scheme between the sensor nodes and the controller node. In case of emergency treatment, the authentication mechanism allows direct access to the sensor nodes.  
 
Li et al. \cite{li2017anonymous} proposed an anonymous and lightweight authentication protocol where a sensor node authenticates with a hub node. In this scheme, the patient's mobile device can be optionally used as a completely trusted forwarder node between the sensor node and the hub. Kompara et al. \cite{kompara2019robust} proposed authentication and key agreement scheme based on Li's scheme \cite{li2017anonymous} that incorporates the session unlinkability property. This scheme also assumes the mobile phone as a trusted entity following \cite{li2017anonymous}. Rehman et al. \cite{rehman2020efficient} extended Kompara's scheme \cite{kompara2019robust} to prevent rogue intermediate node attack, sensor node masquerading attacks, and compromised base station attacks. Likewise, Almuhaideb et al. \cite{almuhaideb2020lightweight} improved the efficiency of kompara's scheme \cite{kompara2019robust} by introducing the concept of re-authentication. In this scheme, a sensor node authenticates with a hub node where a mobile terminal can be used as a forwarder node. Besides, Alzahrani et al. \cite{alzahrani2020provably} offered a lightweight and secure authentication scheme between the sensor node and hub node, where a mobile terminal can also be used as a forwarder. This scheme also assumes the mobile terminal to be trusted.  

Apart from the works discussed above, Jegadeesan et al. \cite{jegadeesan2020epaw} proposed an authentication mechanism between a patient and a doctor that preserves user privacy, data integrity, and non-repudiation property. Further, Mahender and Satish \cite{kumar2020lightweight} introduced an identity-based anonymous authentication and key agreement protocol for WBAN in the cloud-aided environment where a sensor node authenticates with the cloud server. In addition, Chen and Peng \cite{chen2019analysis} proposed an  authentication scheme that mutually authenticates a WBAN client with application provider using asymmetric bilinear pairing. Moreover, Al-Turjman and Alturjman \cite{al2018context} proposed authentication and key agreement mechanism for Wireless Multimedia Medical Sensor Network (WMSN) to support mutual authentication between sensor nodes/smartphones and the medical experts. A healthcare professional collects physiological data from sensor nodes connected to the patient's body in this scheme. Parvez et al. \cite{parvez2019secure} extended this scheme to include a patient's mobile phone in the authentication mechanism. However, this scheme also considers a patient's mobile phone as a trusted entity.

In summary, existing works lack in supporting end-to-end authentication, crucial for the security of health data. Moreover, user authentication mechanisms of WBAN usually assume that the patient's mobile phone gathering data from sensor nodes is trustworthy \cite{li2017anonymous, almuhaideb2020lightweight, alzahrani2020provably}.  In our work, we address these shortcomings and propose a concrete solution that can operate even if the mobile phone is semi-trusted and handle the complete authentication process between a medical expert and a particular sensor node.
\section{System Model} \label{system_model}
\begin{figure}
\centering
\includegraphics[width=\linewidth, height=2.10in]{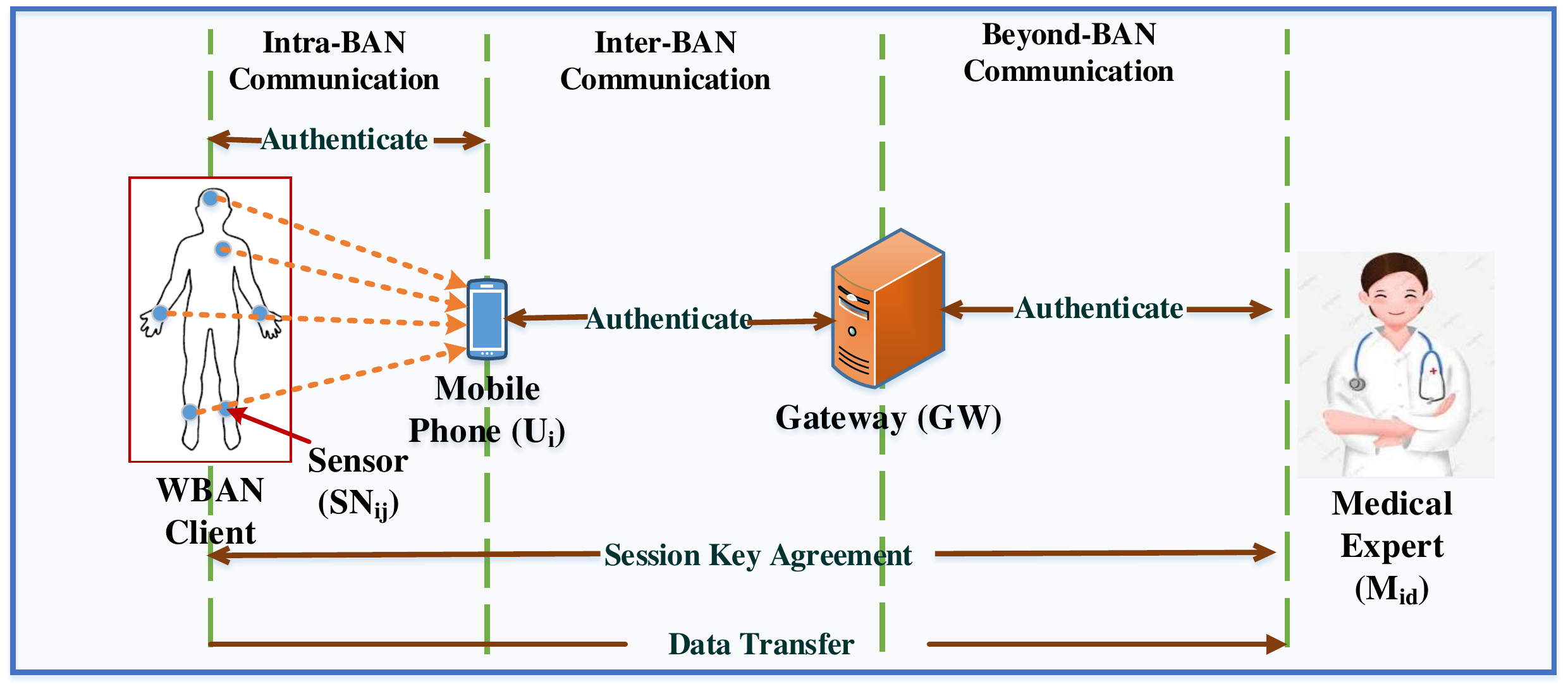}
\caption{System model of the proposed scheme.}
\label{fig:pro_model}
\end{figure}

Figure~\ref{fig:pro_model} presents the system model of our proposed scheme. It comprises \textit{sensors}, \textit{patient's mobile phone}, \textit{gateway server}, and \textit{medical experts}. 

\textit{Sensors} are resource-limited devices attached to the patient's body. They obtain various physiological data and transmit these data with the help of the patient's mobile phone for further processing. We assume that the sensor node works as an honest entity.
	
\textit{Mobile phone} is the patient's portable phone that a patient always carries with them. It accumulates data collected from sensor nodes attached to the patient's body and transmits them for further processing. We assume that the patient's mobile phone is semi-trusted. This situation occurs when a mobile phone is infected by malware. A a semi-trusted entity, the compromised mobile phone accurately follows the protocol but tries to snoop information from the processing \cite{arfaoui2020context}. 
	
\textit{Gateway} is a trustworthy entity managed by a medical organization. It is responsible for registering the patient's mobile phone, the patient's sensor nodes, and medical experts. It also computes secret keys and exchanges them with the corresponding entities using secure communication channels. Besides, gateway takes part in the authentication process between medical professionals and patients.

\textit{Medical Experts} are healthcare providers such as doctors and nurses who periodically monitor the patient's health condition and thus access the patient's health-related information.  

\section{The Proposed Scheme}\label{proposed_scheme}
Our proposed scheme enhances Al-Turjman, and Alturjman's scheme \cite{al2018context} to enable end-to-end user authentication in a realistic WBAN scenario where a patient's mobile phone is semi-trusted. Table~\ref{Table1} lists the symbols used to delineate the proposed scheme.

Our scheme consists of three phases. They are: 
	
\subsection{Registration Phase} 
In this phase, medical experts, patients' mobile phones, and patients' sensor nodes register with a trusted gateway server $GW$.
	
\subsubsection{Medical Expert Registration} 
	The procedure of registering a medical expert consists of the following steps:
	\begin{itemize}
	    \item \textbf{Step 1:} The medical expert selects a unique ID $M_{id}$ and password $PW$ and enters them into their authorized mobile device. This device selects a random number $r_d$, computes $EPW = H(PW \oplus r_d)$  and sends $<M_{id}, EPW>$ to $GW$ using a secure channel. 
		
		\item \textbf{Step 2:} $GW$ computes master keys $K_j$ and $K_l$ for $M_{id}$ \cite{al2018context}. It computes $C= E_{K_j}[M_{id} || ID_{gw}]$ and $N_i= H(M_{id} \oplus EPW \oplus S_{key})$. $GW$ then sends $<H(.), C, N_i, S_{key}, K_j, K_l >$ to $M_{id}$ using a secure communication channel.
\begin{table}
\centering
\caption{List of Notations}
\label{Table1}
\begin{tabular}{|p{1.3cm}|p{6.5cm}|}
\hline
\textbf{Symbol} & \textbf{Description} \\
\hline
$M_{id}$   & Medical expert's ID \\ \hline
$PW$       & Password \\ \hline
$EPW$      & Extended password \\  \hline
$GW$  & Gateway server  \\  \hline
$ID_{gw}$ & Gateway server's ID  \\ \hline
$U_i$  & Mobile phone's ID of $i$th patient \\ \hline
$SN_{j}$ & ID of $j$th sensor node\\ \hline
$S_{key}$ & Secret key between a gateway and a medical expert\\ \hline
$K_j$, $K_l$ &  Master keys between a gateway and a medical expert \\ \hline
$K_{GW-U}$   &  Secret key between a gateway and patient's mobile phone\\ \hline	
$K_{U-SN_{j}}$  &  Secret key between patient's mobile phone and $j$th sensor node \\ \hline
$K_{GW-SN_{j}}$ & Secret key between a gateway and $j$th sensor node \\ \hline
$K_{ssk}$  & Secret session key\\ \hline
$T_i$  & Current timestamp  \\ \hline
$\Delta T_c$  & 	Delay time period \\ \hline
$E_{key}[.]$ & Encryption using $key$ \\ \hline
$D_{key}[.]$ & Decryption using $key$  \\ \hline
$H$(.) & One-way hash function\\  \hline
\end{tabular}
\end{table}
\item \textbf{Step 3:} The medical expert $M_{id}$ stores the received information and $r_d$ in their mobile phone in a secure way. $GW$ also stores $< C, K_j, K_l, H(.) >$ for $M_{id}$.

	\end{itemize}
\subsubsection{Patient's Mobile Device Registration}
	The gateway selects a unique ID $U_i$ for a patient's mobile phone and computes $K_{GW-U} = H(U_i \oplus ID_{gw})$. It then securely shares $<U_i,K_{GW-U}>$ with the patient's mobile phone. $GW$ also stores $K_{GW-U}$ for $U_i$.
	
	\subsubsection{Sensor Registration}
	The gateway $GW$ assigns a unique ID $SN_{j}$ to the $j$th sensor node connected to $U_i$ and computes $K_{U-SN_{j}} = H(U_i \oplus SN_{j})$ and $K_{GW-SN_{j}} = H(ID_{gw} \oplus SN_{j})$. It sends $<U_i,SN_{j}, K_{U-SN_{j}}, K_{GW-SN_{j}}, H(.)>$ to the sensor node via a secure channel. Besides, it securely shares $<SN_{j}, K_{U-SN_{j}}>$ to the mobile phone $U_i$. $GW$ also stores $K_{GW-SN_{j}}$ and $K_{U-SN_{j}}$ for $SN_{j}$.	
	
\subsection{Authentication Phase}
In this phase, a medical expert, a patient's mobile phone, and a specific sensor node authenticate each other through mutual authentication. At the end of this phase, a medical expert and a sensor node establish a unique session key to continue their future communications. The required steps of this phase are as follows:
\begin{itemize}
	\item \textbf{Step 1:}
		The medical expert enters $M_{id}$ and $PW$ to their  authorized mobile device to log in to the system. This device calculates ${N^*_i} = H(M_{id} \oplus H(PW \oplus r_d) \oplus S_{key})$ with the supplied $M_{id}$ and $PW$. If $N^*_i = N_i$, the medical expert can proceed for further computations. This step prevents a wicked person to use the device allocated to a honest medical professional. The medical expert calculates $H(M_{id})$ and $CID_i = E_{K_l}[H(M_{id} ) || M || U_i || SN_{j} || C || T_1 ] $ where $M$ is a randomly selected nonce.  The medical expert transmits $<CID_i , C, T_1>$ to $GW$ using a public channel.
	
	\item \textbf{Step 2:}
		The gateway $GW$ checks into its database for $C$ and fetches corresponding $K_j$ and $K_l$. It computes $D_{K_l}[CID_i]$ and excerpts $H(M'_{id}), M, U_i, SN_{j}$, $C^*$ and ${T^*_1}$ from $CID_i$. If $T_1 = {T^*_1}$ and  $T_2 - T_1 \leq \Delta T_c$, $GW$ continues further processing where $T_2$ is the time when $GW$ receives $<CID_i , C, T_1>$. This test ensures $<CID_i , C, T_1>$ is received within a permitted time period $\Delta T_c$. Besides, $GW$ computes $D_{K_j}[C]$, extracts	${M^*_{id}}, ID^*_{gw}$ from $C$ and computes $H(M^*_{id})$. If $H(M^*_{id}) = H(M'_{id})$ and $ID^*_{gw} = ID_{gw}$ then $GW$ computes $X = E_{K_{GW-SN_{j}}}[M_{id} || M]$ and $V_i = E_{K_{GW-U}}[U_i || SN_{j} || X || T_3]$ and sends $<V_i, T_3>$ to the patient's mobile phone $U_i$.
	
	\item \textbf{Step 3:}
		The mobile phone $U_i$ computes $D_{K_{GW-U}}[V_i]$ and excerpts ${U^*_i},$ ${SN_{j}}$, $X$ and ${T_3}^*$ from $V_i$. It checks for $T^*_3 = T_3$ and $T_4 - T_3 \leq \Delta T_c$ where $T_4$ is the time when $U_i$ receives $<V_i, T_3>$. If both conditions are satisfied and ${U^*_i} = U_i$ then the mobile phone computes $ V'_i = E_{K_{U-SN_{j}}}[X || U_i || SN_{j} || T_5]$ and sends $< V'_i , T_5>$ to $SN_{j}$ through a public channel.
		
	\item \textbf{Step 4:}
		The sensor node $SN_{j}$ computes $D_{K_{U-SN_{j}}}[V'_i]$ and retrieves $X, {U_i}, {SN^*_{j}}$ and ${T^*_5}$ from $V'_i$. Besides, it computes $D_{K_{GW-SN_{j}}}[X]$ and extracts ${M^*_{id}}$ and $M^{*}$. If $T^*_5 = T_5$ and $T_6 - T_5 \leq \Delta T_c$  where $T_6$ is the time when $SN_{j}$ receives $< V'_i , T_5>$, the sensor performs subsequent computations.  If ${SN^*_{j}} = SN_{j}$ the sensor calculates $K_{ssk} = H(M^*_{id} \oplus SN^*_{j} \oplus M^*)$. Alongside, it computes $L= E_{K_{ssk}}[SN^*_{j} || M^*_{id} || T_7]		$ and sends $<L, T_7>$ to the medical expert over the public channel.
		
	\item \textbf{Step 5:}
		The medical expert also computes $K_{ssk} = H(M_{id} \oplus S_{N_j} \oplus M)$ using information stored in their mobile device. $M_{id}$ also computes $D_{K_{ssk}}[L]$ and retrieves ${SN^*_{j}}, {M^*_{id}}$ and  ${T^*_7}$. If $T^*_7 = T_7$ and $T_8 - T_7 \leq \Delta T_c$ the medical expert continues future computations where $T_8$ is the time when $M_{id}$ receives $< L, T_7>$. If ${SN^*_{j}} = SN_{j}$ and ${M^*_{id}}  = M_{id}$,  then the medical expert is confirmed that the same secret key $K_{ssk}$ is set up between $M_{id}$ and $SN_{j}$.
\end{itemize}
    
    \subsection{Password Update Phase}
	 	To update the password, the medical expert must log in to the system. The necessary steps are as follows:
	 	\begin{itemize}
	 	\item\textbf{Step 1:}
	 	The medical professional enters $M_{id}$ and $PW$ to their assigned mobile device. The device computes ${N_i}^* = H(M_{id} \oplus H(PW \oplus r_d) \oplus S_{key})$ and compares $N^*_i = N_i$. If the comparison is true, the medical expert can proceed further computations.
	 	\item\textbf{Step 2:}
	 	 The medical expert $M_{id}$ enters a new password $PW^{new}$. The device again chooses a random number ${r^{new}_d}$, computes $EPW^{new} = H(PW^{new} \oplus {r^{new}_d})$ and sends 
	 	$<M_{id}, EPW^{new}>$ to $GW$ through a secure communication channel.\vspace{0.5em}
	 	
	 	\item\textbf{Step 3:}
	 	 $GW$ computes $N^{new}_i= H(M_{id} \oplus EPW^{new} \oplus S_{key})$ and sends $N^{new}_i$ to $M_{id}$ using a secure channel. The device replaces $N_i$ and $r_d$ with $N^{new}_i$ and ${r^{new}_d}$ in its memory.
         \end{itemize}
\section{Security Analysis}\label{security_analysis}
We first present the security properties preserved when the patient's mobile phone works genuinely. Subsequently, we discuss the resiliency of the proposed scheme when the patient's mobile phone is compromised.

\subsection{Security Analysis when Patient's Mobile Phone is Trusted}
\begin{itemize}
	\item\textbf{Mutual Authentication:}
	A medical expert $M_{id}$ and a sensor $SN_{j}$ connected to a patient authenticate each other to set up a secure communication. During registration phase, $GW$ transmits $C$=$E_{K_j}[M_{id} || ID_{gw}]$ to the medical expert. In the authentication phase, $M_{id}$ sends the same $C$ along with $CID_i$ to $GW$. The gateway computes $D_{K_j}[C] = D_{K_j}[{M^*_{id}} || {ID^*_{gw}}]$ and $D_{K_l}[CID_i] = [H(M'_{id}) || M || U_i || SN_{j} || T_1]$. It authenticates $M_{id}$ when $H(M^*_{id}) = H(M'_{id})$ and $ID^*_{gw} = ID_{gw}$. Furthermore, $GW$ generates $X = E_{K_{GW-SN_{j}}}[M_{id} || M]$ and then $V_i = E_{K_{GW-U}}[U_i || SN_{j} || X || T_3]$. As $K_{GW-U}$ is shared between $GW$ and $U_i$, the patient's mobile phone can decrypt $V_i$. $U_i$ computes $V'_i$ using $K_{U-SN_{j}}$. As $K_{U-SN_{j}}$ is a secret between $U_i$ and $SN_{j}$, only the sensor node can decrypt $V'_i$ and retrieve $U_i$, $SN_{j}$ and $X$. The sensor node further decrypts $X$ using $K_{GW-SN_{j}}$ and obtains $M_{id}, M$ to generate $K_{ssk}$. It also computes $L= E_{K_{ssk}}[SN_{j} || M_{id} || T_7]$ and sends $< L, T_7>$ to the medical expert. $M_{id}$ computes $K_{ssk}$ using $M_{id}$, $SN_{j}$, and $M$ available to its storage. $M_{id}$ then decrypts $L$ using $K_{ssk}$ and obtains the identifies of the medical expert and the sensor node. If these identities match with those parameters sent through $CID_i$, the medical expert is sure that same $K_{ssk}$ is generated between $M_{id}$ and $SN_{j}$. Hence, our scheme ensures mutual authentication.                                                          
	\item\textbf{Unique Secret Key Generation:}
	After successful authentication, a sensor node and a medical expert share a secure session key. This session key is calculated as $K_{ssk}$ = $H(M_{id} || M || SN_{j})$. Since the medical expert selects a new random nonce $M$ in every session, a unique session key is created for each new data transfer operation between $M_{id}$ and $SN_{j}$.
	
	\item\textbf{User Masquerading Attack:}
	An adversary can capture $<CID_i, C, T_1>$ as this message is transmitted through a public channel. They may try to alter the message and introduce a new message $<{CID^{new}_i}, C, {T_1} >$ in the channel where ${CID^{new}_i}$ is constructed using ${M^*_{id}}$, $M^{*}$, ${U^*_i}$, ${SN^*_{j}}$ and $T_1$ selected by the adversary. As $K_l$ (distributed between $GW$ and $M_{id}$) is not known to the adversary, they cannot produce ${CID^{new}_i}$ in a correct form that can be decrypted successfully by $GW$ using $K_l$. Similarly, an adversary cannot counterfeit $C$ as $K_j$ is not known. An adversary also cannot forge $V_i$, ${V'_i}$ due to the lack of access to $K_{GW-U}$ and $K_{U-SN_{j}}$, respectively. Moreover, the adversary cannot regenerate $X$ due to the lack of access to   $K_{GW-SN_j}$. Besides, they cannot reproduce $L$ as $M_{id}$, $SN_{j}$ and $M$ are unknown. Thus, masquerading a user is not possible.

	\item\textbf{Secret Gateway Guessing Attack:}
	Our scheme utilizes six different keys such as $K_j$, $K_l$, $S_{key}$, $K_{GW-U}$, $K_{GW-SN_{j}}$ and $K_{U-SN_{j}}$. $GW$ shares these keys with different entities in a secure way. Moreover, an adversary cannot obtain these keys from $GW$ as it is a trusted entity. Besides, our scheme exchanges the identities of medical experts, patient's mobile phones, sensor nodes, and $H(.)$ in a secure way. Therefore, the adversary cannot guess or reproduce $K_{GW-U}$, $K_{GW-SN_{j}}$ and $K_{U-SN_{j}}$. Hence, secret gateway guessing attacks are not possible. In addition, the adversary is not able to compute the session key as $H(.)$, $M_{id}$, $M$ and $SN_{j}$ are hidden. 
	
	\item\textbf{Replay Attack:}
	An adversary cannot utilize previous obsolete messages $<CID_i, C, T_1>$, $<V_i, T_3>$, $< V'_i, T_5>$, and $< L, T_7>$ to access the system. They can alter the timestamp component of these messages only. Besides, $CID_i$, $V_i$, $V'_i$, and $L$ also include the timestamp $T_i$. An adversary cannot change $T_i$ in these messages due to not having access to the necessary keys. Hence, comparing the timestamp obtained from $CID_i$, $V_i$, $V'_i$, and $L$ with the timestamp component modified by the adversary in the message request will never be successful. Moreover, each entity of the WBAN also ensures that messages are received within a pre-defined time frame $\Delta T_c$. Therefore, our scheme is resilient to replay attacks.
	
	\item\textbf{Man-in-the-middle attack:}
	In this attack, an adversary can snoop and possibly alter the messages transmitted through the communication channel without informing the communicating parties. Since attackers do not have access to the secret keys $K_j$, $K_l$, $K_{GW-U}$, $K_{GW-SN_{j}}$, $K_{U-SN_{j}}$ and $K_{ssk}$, they can not recover the original message by eavesdropping or can not reconstruct a new message that decrypts successfully. Therefore, a man-in-the-middle attack is not possible.
	
	\item\textbf{User Anonymity:}
	The proposed scheme hides the identities of patients' mobile phones, sensor nodes, and medical experts from unauthorized parties. All this information is stored in encrypted form in the messages exchanged during the authentication process. Since the adversaries do not have access to the required keys to decrypt these messages, they cannot gain any information regarding the identities of patients' mobile phones, sensor nodes, and medical experts. Thus our scheme ensures the anonymity of patients and medical professionals.
	
	\item \textbf{Forward and Backward Secrecy:}
	The proposed scheme ensures that the compromise of a  session key does not hamper the secrecy of previous and future sessions. In our scheme, $K_{ssk}$ is generated as $H(M_{id} \oplus SN_{j} \oplus M)$ and for the use of one-way hash function $H(.)$, it is not possible to extract $M_{id}$, $SN_{j}$, and $M$. Moreover, the identities of sensors, mobile phones, and medical experts are always transmitted in an encrypted form. Due to not having access to the decryption keys, an adversary cannot retrieve that information. In addition, $M$ changes in every session to generate a unique key. Therefore, it is not possible to construct any previous and future session keys when a session key is exposed.
    \end{itemize}
    
\subsection{Patient's Mobile Device is Compromised}
    The proposed scheme prevents false authentication in case of patient's mobile phone is compromised for example through malware attacks. The mobile device $U_i$ receives $<V_i, T_3>$ from $GW$ where $V_i = E_{K_{GW-U}}[U_i || SN_{j} || X || T_3]$ and $X =E_{K_{GW-SN_{j}}}[M_{id}$ $ || M]$. It decrypts $V_i$ using $K_{GW-U}$ and obtains $X$. $U_i$ is unable to decrypt $X$ due to lack of access to $K_{GW-SN_{j}}$. Therefore, it cannot obtain $M_{id}$ and $M$ which are necessary to generate the session key. Moreover, the mobile phone delivers $< V'_i , T_5>$ to the sensor node where $V'_i = E_{K_{U-SN_{j}}}[X || U_i || SN_{j} || T_5]$, and $SN_{j}$ decrypts $X$ using $K_{GW-SN_{j}}$ and forms the session-key $ssk$.
    
    As $U_i$ is compromised, an adversary can obtain the secret key $K_{GW-U}$ for a particular patient's mobile phone and the secret key $K_{U-{SN_j}}$ of the sensor nodes associated with that mobile device. As the adversary does not have access to $K_{GW-{SN_j}}$, they cannot decrypt $X$. Hence, it is not possible for an adversary to obtain $M_{id}$ and $M$ required for a session key. Also, the adversary cannot alter $X$ without the possession of $K_{GW-{SN_j}}$. 
    
    An adversary can reconstruct $V_i$ and $V_{i'}$ for $U_i$ and arbitrary $SN_j$ as they possess $K_{GW-U}$ and $K_{U-{SN_j}}$ (obtained from compromised patients' mobile phones). In the worst case, they can incorporate a $X$ captured from previous sessions involving the same $U_i$, $GW$ and $SN_{j}$ with the reconstructed $V_i$ and $V_{i'}$. Besides, they can manipulate the timestamp component in $V_i$ and $V_i'$. Thus an adversary can replay $V_i$ and $V_i'$ in the channel that a medical expert does not initiate. In this case, when a $L$ is reached to the medical experts, they can identify the false attempt to establish a session, and $L$ may be reached after the pre-defined time interval. Hence, the adversary cannot get any advantages by replaying $V_i$ and $V_i'$.
    
    Due to access of $K_{GW-U}$ and $K_{U-{SN_j}}$, an adversary can obtain information about a particular patient and sensor nodes associated with that patient. Possession of $K_{GW-U}$ and $K_{U-{SN_j}}$ does not help to identify the medical expert. An adversary also cannot construct the secret keys of other entities with the help of the leaked identities as they do not have access to $H(.)$.
\section{Protocol Analysis using BAN Logic}\label{protocol_analysis}
We use BAN logic \cite{burrows1989logic} to verify the validity of the proposed scheme in generating secret session keys. Table \ref{Table2} presents a brief description of the notations used for the BAN logic \cite{burrows1989logic}.

\begin{table}
\centering
\caption{List of Notations used in BAN Logic}
\label{Table2}
\scalebox{0.90}{
\begin{tabular}{|p{1.2cm}|p{2.5cm}||p{1.2cm}|p{2.5cm}|}
\hline
\textbf{Notation} & \textbf{Narration} & \textbf{Notation} & \textbf{Narration} \\ \hline
\hline
$\{Y\}_K$ & $Y$ is encrypted by $K$ & $R\Rightarrow Y$ & $R$ controls $Y$ \\ \hline
$R\triangleleft \{Y\}_K$ & $R$ sees $\{Y\}_K$ & $\#(Y)$ & $Y$ is fresh \\ \hline
$R\mid\sim Y$ & $R$ said $Y$ & $R\xleftrightarrow{K}S$ & $R$ and $S$ shares $K$ \\
\hline
$R\mid\equiv Y$ & $R$ believes $Y$ & & \\ \hline
\end{tabular}}
\end{table}

We need to satisfy the following goals to confirm the security of the proposed scheme:

\noindent\textbullet Goal 1: $MD\mid\equiv SN_j\mid\equiv (MD\xleftrightarrow{K_{ssk}}SN_j)$

\noindent\textbullet Goal 2: $SN_j\mid\equiv MD\mid\equiv (MD\xleftrightarrow{K_{ssk}}SN_j)$

\noindent\textbullet Goal 3: $MD\mid\equiv  (MD\xleftrightarrow{K_{ssk}}SN_j)$

\noindent\textbullet Goal 4: $SN_j\mid\equiv (MD\xleftrightarrow{K_{ssk}}SN_j)$

We use the BAN logic rules,  idealized messages, and assumptions to prove that the proposed scheme satisfies the security goals. Table \ref{Table3} shows the BAN logic rules \cite{burrows1989logic} used in our analysis.
\begin{table}
\centering
\caption{BAN Logic Rules}
\begin{tabular}{|p{2.5cm}|p{5cm}|}
\hline
\textbf{Rule} & \textbf{Narration} \\
\hline
$\frac{R\mid\equiv R\xleftrightarrow{K}S, R \triangleleft \{Y\}_K}{R\mid\equiv S\mid\sim Y}$ & $R_1$ (Message-meaning rule): If $R$ believes that $R$ shares $K$ with $S$ and $R$ observes $Y$ encrypted with $K$, $R$ trusts $S$ said $Y$ \\ \hline
$\frac{R\mid\equiv \#(Y), R\mid\equiv S \mid\sim Y}{R \mid\equiv S \mid\equiv Y}$ & $R_2$ (Nonce-verification rule): If $R$ believes that $Y$ is new and $R$ believes $S$ uttered $Y$, $R$ believes $S$ trusts $Y$ \\
\hline
$\frac{R\mid\equiv \#(Y)}{R\mid\equiv \#(Y, Z)}$ & $R_3$ (Freshness-conjunction rule): If $R$ trusts that $Y$ is new, $R$ admits $(Y, Z)$ is fresh \\ 
\hline
$\frac{R\mid\equiv S \Rightarrow Y, R\mid\equiv S \mid\equiv Y}{R \mid\equiv Y}$ & $R_4$ (Jurisdiction rule): If $R$ believes $S$ controls $Y$ and $R$ believes $S$ trusts $Y$, $R$ trusts $Y$ \\
\hline
\end{tabular}
\label{Table3}
\end{table}

The idealized form of the transmitted messages are as follows:

\small\noindent\textbullet $M_1: MD \rightarrow GW: \{H(M_{id}), M, U_i, SN_j, C, T_1\}_{k_l}$
 
\noindent\textbullet $M_2: GW\rightarrow U_i: \{U_i, SN_j, \{M_{id},M\}_{K_{GW-SN_j}},T_3\}_{K_{GW-U}}$
 
\noindent\textbullet $M_3: U_i\rightarrow SN_j:\{\{M_{id}, M\}_{K_{GW-SN_j}}, U_i, SN_j, T_5\}_{K_{U-SN_j}}$

\noindent\textbullet $M_4:SN_j\rightarrow MD: \{SN_j, M_{id}, T_7\}_{K_{ssk}}$

We extract the following initial assumptions from the protocol messages:

\noindent \textbullet $P_1: MD\mid\equiv MD\xleftrightarrow{K_{l}}GW$

\noindent \textbullet $P_2: GW\mid\equiv MD\xleftrightarrow{K_{l}}GW$

\noindent \textbullet $P_3: MD\mid\equiv MD\xleftrightarrow{K_{j}}GW$

\noindent \textbullet $P_4: GW\mid\equiv MD\xleftrightarrow{K_{j}}GW$

\noindent \textbullet $P_5: GW\mid\equiv \# (T_1)$

\noindent \textbullet $P_6: GW\mid\equiv \# (M)$

\noindent \textbullet $P_7: GW\mid\equiv U_i\xleftrightarrow{K_{GW-U}}GW$

\noindent \textbullet $P_8: U_i\mid\equiv U_i\xleftrightarrow{K_{GW-U}}GW$

\noindent \textbullet $P_9: U_i\mid\equiv \# (T_3)$

\noindent \textbullet $P_{10}: U_i\mid\equiv SN_j\xleftrightarrow{K_{U-SN_j}}U_i$

\noindent \textbullet $P_{11}: SN_j\mid\equiv SN_j\xleftrightarrow{K_{U-SN_j}}U_i$

\noindent \textbullet $P_{12}: SN_j\mid\equiv \# (T_5)$

\noindent \textbullet $P_{13}: GW\mid\equiv SN_j\xleftrightarrow{K_{GW-SN_j}}GW$

\noindent \textbullet $P_{14}: SN_j\mid\equiv SN_j\xleftrightarrow{K_{GW-SN_j}}GW$

\noindent \textbullet $P_{15}: MD\mid\equiv \# (T_7)$

\noindent \textbullet $P_{16}: SN_j\mid\equiv SN_j\xleftrightarrow{K_{ssk}}MD$

\noindent \textbullet $P_{17}: MD\mid\equiv SN_j\xleftrightarrow{K_{ssk}}MD$


\noindent \textbullet $P_{18}: MD\mid\equiv SN_j\Rightarrow{(MD\xleftrightarrow{K_{ssk}}SN_j)}$

\noindent \textbullet $P_{19}: SN_j\mid\equiv MD\Rightarrow{(MD\xleftrightarrow{K_{ssk}}SN_j)}$

\noindent \textbullet $P_{20}:SN_j\mid\equiv\#(M)$

\vspace{1em}From $M_1$, we get

\noindent\textbullet $V_1:GW\triangleleft\{H(M_{id}), M, U_i, SN_j, C, T_1\}_{k_l}$

From $P_2$ and $V_1$ using $R_1$ we get 

\noindent\textbullet $V_2:GW\mid\equiv MD\mid\sim \{H(M_{id}), M, U_i, SN_j, C, T_1\}$

From $P_5$ and $P_6$ using $R_3$ we get

\noindent\textbullet $V_3:GW\mid\equiv \# \{H(M_{id}), M, U_i, SN_j, C, T_1\}$

From $V_2$ and $V_3$ using $R_2$, we get

\noindent\textbullet $V_4:GW\mid\equiv MD\mid\equiv \{H(M_{id}), M, U_i, SN_j, C, T_1\}$
\\

From $M_2$, we get

\noindent\textbullet$V_5:U_i\triangleleft\{U_i, SN_j, \{M_{id},M\}_{K_{GW-SN_j}},T_3\}_{K_{GW-U}}$

From $P_8$ and $V_5$ using $R_1$, we get

\noindent\textbullet$V_6:U_i\mid\equiv GW\mid\sim\{U_i, SN_j, \{M_{id},M\}_{K_{GW-SN_j}},T_3\}$

From $P_9$ and $R_3$ we get

\noindent\textbullet$V_7:U_i\mid\equiv\#\{U_i, SN_j, \{M_{id},M\}_{K_{GW-SN_j}},T_3\}$

From $V_6$ and $V_7$ using $R_2$, we get

\noindent\textbullet$V_8:U_i\mid\equiv GW\mid\equiv\{U_i, SN_j, \{M_{id},M\}_{K_{GW-SN_j}},T_3\}$
\\

From $M_3$, we get

\noindent\textbullet$V_9:SN_j\triangleleft\{\{M_{id}, M\}_{K_{GW-SN_j}}, U_i, SN_j, T_5\}_{K_{U-SN_j}}$

From $P_{11}$ and $V_9$ using $R_1$ we get,

\noindent\textbullet$V_{10}:SN_j\mid\equiv U_i\mid\sim \{\{M_{id}, M\}_{K_{GW-SN_j}}, U_i, SN_j, T_5\}$

From $P_{12}$ and $R_3$ we get,

\noindent\textbullet$V_{11}:SN_j\mid\equiv \# \{\{M_{id}, M\}_{K_{GW-SN_j}}, U_i, SN_j, T_5\}$

From $V_{10}$ and $V_{11}$, using $R_2$ we get,

\noindent\textbullet$V_{12}:SN_j\mid\equiv U_i\mid\equiv \{\{M_{id}, M\}_{K_{GW-SN_j}}, U_i, SN_j, T_5\}$
\\

From $M_4$, we get

\noindent\textbullet$V_{13}:MD\triangleleft\{SN_j, M_{id}, T_7\}_{K_{ssk}}$

From $P_{17}$ and $V_{13}$ using $R_1$ we get,

\noindent\textbullet$V_{14}:MD\mid\equiv SN_j\mid\sim \{SN_j, M_{id}, T_7\}$

From $P_{15}$ using $R_3$ we get,

\noindent\textbullet$V_{15}:MD\mid\equiv \# \{SN_j, M_{id}, T_7\}$

From $V_{14}$ and $V_{15}$ using $R_2$ we get,

\noindent\textbullet$V_{16}:MD\mid\equiv SN_j\mid\equiv \{SN_j, M_{id}, T_7\}$
\\

From $V_{12}$ we get

\noindent\textbullet$V_{17}:SN_j\triangleleft\{M_{id}, M\}_{K_{GW-SN_j}}$

From $P_{14}$ and $V_{17}$ using $R_1$ we get

\noindent\textbullet$V_{18}:SN_j\mid\equiv GW\mid\sim\{M_{id}, M\}$

From $P_{20}$ and $R_3$ we get

\noindent\textbullet$V_{19}:SN_j\mid\equiv \mid\#\{M_{id}, M\}$

From $V_{18}$ and $V_{19}$ using $R_2$ we get

\noindent\textbullet$V_{20}:SN_j\mid\equiv GW \mid\equiv\{M_{id}, M\}$
\\

As $K_{ssk}$ = $H(M_{id} \oplus SN_j \oplus M)$ and combining $V_{20}$, $V_{12}$, $V_8$ and $V_4$ we get 

\noindent\textbullet$V_{21}:SN_j\mid\equiv MD \mid\equiv(MD\xleftrightarrow{K_{ssk}}SN_j)$ (Goal 2)

As $K_{ssk}$ = $H(M_{id} \oplus SN_j \oplus M)$, from $V_{16}$ we get

\noindent\textbullet$V_{22}:MD\mid\equiv SN_j\mid\equiv(MD\xleftrightarrow{K_{ssk}}SN_j)$ (Goal 1)

From $V_{21}$ and $P_{19}$ using $R_4$ we get

\noindent\textbullet$V_{23}:SN_j\mid\equiv(MD\xleftrightarrow{K_{ssk}}SN_j)$ (Goal 4) 

From $V_{22}$ and $P_{18}$ using $R_4$ we get

\noindent\textbullet$V_{24}:MD\mid\equiv(MD\xleftrightarrow{K_{ssk}}SN_j)$ (Goal 3) 
\section{Comparative Study}\label{comparative_study}
In this section, we present a comparison of the proposed scheme with the other related works: the schemes of Abiramy and Sudha \cite{abiramy2019secure}, Li et al. \cite{li2017anonymous} and Al-Turjman and Alturjman \cite{al2018context} in respect to computation, communication, and security features.
\begin{table}
\centering
\caption{Comparison based on Security Features}
\label{Table4}
\scalebox{0.80}{
\begin{tabular}{|p{4cm}|p{1.2cm}|p{1.45cm}|p{1cm}|p{1cm}|}
\hline
\textbf{Features} & \textbf{Abiramy and Sudha \cite{abiramy2019secure}} & \textbf{Al-Turjman and Alturjman \cite{al2018context}} & \textbf{Li et al. \cite{li2017anonymous}} & \textbf{Proposed scheme} \\
\hline
Mutual authentication&\checkmark &\checkmark&\checkmark&\checkmark\\
\hline
User anonymity&$\times$&\checkmark&\checkmark&\checkmark \\
\hline
Resilient to semi-trusted mobile device& $\times$ &--& $\times$&\checkmark\\
\hline
End-to-end authentication & $\times$ & $\times$ & $\times$ & \checkmark \\
\hline
User masquerading attack& 
   --&\checkmark&\checkmark&\checkmark \\
\hline
Replay attack&
  --&\checkmark&\checkmark&\checkmark\\
  \hline
Man-in-the-middle attack& 
  --& \checkmark&\checkmark&\checkmark \\
\hline
Secret gateway guessing attack & -- & \checkmark &\checkmark&\checkmark \\
\hline
Forward and backward secrecy & forward secrecy only &\checkmark & \checkmark&\checkmark\\
\hline
\end{tabular}}
\begin{tablenotes}
\item \footnotesize{\checkmark: a scheme conserves a feature; $\times$: a scheme does not conserve a feature; $-$: a scheme unresponsive about a feature.}
\end{tablenotes}
\end{table}
\subsection{Comparison in respect to Security Features}

A comparison of our scheme with the existing techniques based on security features is shown in Table \ref{Table4}.  It is apparent from the table that our scheme provides better security compared to the other schemes. Notably, none of the current schemes raise protection against the semi-trusted patient's mobile phone. Abiramy and Sudha \cite{abiramy2019secure} and Li et al. \cite{li2017anonymous} assumed that patient's mobile phone is their scheme is fully trusted. Besides, existing schemes do not provide end-to-end authentication. Furthermore, Abiramy and Sudha \cite{abiramy2019secure} does not achieve user anonymity and provides forward secrecy only.

\subsection{Comparison in respect to Computation Cost}
The computation cost is measured as the total time required to perform mutual authentication. We use $T_H$, $T_{XOR}$, $T_{ENC}$, $T_M$, and $T_{EXP}$ to denote the time required to compute hash function, ex-or operation, symmetric key encryption/decryption, Elliptic Curve Cryptography (ECC) point multiplication, and exponentiation operation, respectively. Table \ref{Table5} shows the computation cost of each entity in our proposed scheme. In total the computation cost of the proposed scheme is $6T_H + 11T_{ENC} + 7T_{XOR} \approx 6T_H + 11T_{ENC}$ (Here to mention that $T_{XOR}$ is negligible compared to the other costs). Table \ref{Table6} presents a comparison of our scheme with the existing schemes in respect to computation cost. Our scheme attains higher computation time compared to the Al-Turjman and Alturjman scheme \cite{al2018context} due to the explicit inclusion of the patient's mobile phone in the authentication process and preventing crucial information for session key generation from the semi-trusted mobile phone. These two features are missing in other low-cost related works.
\begin{table}
\centering
\caption{Computation Cost of Our Scheme}
\label{Table5}
\begin{tabular}{|p{2.5cm}|p{4cm}|}
\hline
\textbf{Node} & \textbf{Computation cost} \\
\hline
Medical expert & 4$T_H$ + 5$T_{XOR}$ + 2$T_{ENC}$\\
\hline
Gateway & 4$T_{ENC}$ + $T_H$ \\
\hline
Mobile device & 2$T_{ENC}$\\
\hline
Sensor & 3$T_{ENC}$ + $T_H$ + 2$T_{XOR}$\\
\hline
\end{tabular}
\end{table}
	
\begin{table}
\centering
\caption{Comparison based on Computation Cost}
\label{Table6}
\scalebox{0.95}{
\begin{tabular}{|p{3cm}|p{5cm}|}
\hline
\textbf{Scheme} & \textbf{Computation cost}\\
\hline
Abiramy and Sudha \cite{abiramy2019secure} & 9$T_M$ + 2$T_{EXP}$ + 3$T_H$\\
\hline
Al-Turjman and Alturjman \cite{al2018context} &  6$T_H$ + 7$T_{ENC}$ + 2$T_{XOR}$ $\approx$ 6$T_H$ + 7$T_{ENC}$\\
\hline
Li et al. \cite{li2017anonymous} & 8$T_H$ + 17$T_{XOR}$ $\approx$ 8$T_H$\\
\hline
Proposed scheme & 6$T_H$ + 11$T_{ENC}$ + 7$T_{XOR}$ $\approx$ 6$T_H$ + 11$T_{ENC}$\\
\hline
\end{tabular}}
\end{table}

\subsection{Comparison in respect to Communication Cost}
Table \ref{Table7} presents the communication cost of our proposed scheme. We assume $|M_{id}|$ = $|U_i|$ = $|SN_j|$ = 32 bits \cite{saeed2018remote}, $|T_i|$ = 32 bits \cite{jegadeesan2020epaw} and $|M|$ = 64 bits \cite{li2018secure}. Besides, we use SHA-1 hash algorithm \cite{SHA1} for hash function and AES-128 \cite{AES2001} for symmetric key encryption. In $M_{id}\rightarrow GW$, the medical expert sends the tuple $<CID_i , C, T_1>$ where the size of $C$ is 128 bits and $CID_i$ is (160 + 64 + 32 + 32 + 128 + 32) = 448 bits. AES-128 divides $CID_i$ into 4 blocks that consume 4 $\times$ 128 = 512 bits. Therefore, $<CID_i , C, T_1>$ incurs communication cost of (128 + 512 + 32) = 672 bits. In  the next step, $GW$ sends $<V_i, T_3>$ to $U_i$ where the size of $V_i$ is (128 + 32 + 32 + 32) = 224 bits $\approx$ 2 AES blocks $\approx$ 256 bits. Hence, $<V_i, T_3>$ imposes (256 + 32) = 288 bits of communication overhead. For the similar reasoning, $U_i$ sends to $SN_j$ $<V'_i, T_5>$ of size 288 bits. Finally, $SN_j$ sends $<L, T_7>$ of size 160 bits to $M_{id}$ to complete the authentication process. Table \ref{Table8} presents a comparison of the proposed scheme with the existing related works in respect to the communication cost. We observe that
our scheme achieves lesser communication cost compared to Li et al. \cite{li2017anonymous} scheme and greater communication cost compared to Al-Turjman and Alturjman \cite{al2018context} scheme. Al-Turjman and Alturjman \cite{al2018context} scheme does not include patient's mobile phone in the authentication process and therefore, eliminates message communication with the mobile phone. Moreover, this scheme does not provide protection against semi-trusted patient's mobile phone. Here to mention that, although our scheme generates higher computation cost compared to the Li et al. \cite{li2017anonymous} scheme, our scheme achieves significant reduction in communication cost. Li et al. \cite{li2017anonymous} scheme mainly uses hash function to ensure lightweight computation and hence, each entity transmitted over the communication channel is 160 bits and they are sent in clear text, that ultimately increases communication cost.

\begin{table}
\centering
\caption{Communication Cost of Our Scheme}
\label{Table7}
\scalebox{0.95}{
\begin{tabular}{|p{4cm}|p{4cm}|}
\hline
\textbf{Communication between nodes} & \textbf{Communication cost}\\
\hline
$M_{id}\rightarrow GW$ & 672 bits \\
\hline
$GW\rightarrow U_i$ & 288 bits \\
\hline
$U_i \rightarrow SN_j$ & 288 bits\\
\hline
$SN_j \rightarrow M_{id}$ & 160 bits \\
\hline
Total & 1408 bits\\
\hline
\end{tabular}}
\end{table}

\begin{table}
\centering
\caption{Comparison based on Communication Cost}
\scalebox{0.95}{
\begin{tabular}{|p{4cm}|p{4cm}|}
\hline
\textbf{Scheme} & \textbf{Communication Cost}\\
\hline
Abiramy and Sudha \cite{abiramy2019secure} & --\\
\hline
Al-Turjman and Alturjman \cite{al2018context} &  864 bits\\
\hline
Li et al. \cite{li2017anonymous} & 2656 bits\\
\hline
Proposed scheme & 1408 bits \\
\hline
\end{tabular}}
\label{Table8}
\end{table}
\section{Practical Impact Study}\label{practical_impact}
We implemented the proposed scheme using Network Simulator 3 (NS-3) \cite{riley2010ns} and studied the impact of our scheme on various network performance parameters such as \textit{throughput} (in bytes per second) and \textit{end-to-end delay} (in second). Table \ref{Table9} presents the parameters for NS-3 simulation. 

\begin{table}
\centering
\caption{Parameters used in Simulation}
\label{Table9}
\begin{tabular}{|p{3.5cm}|p{4cm}|}
\hline
\textbf{Parameter} & \textbf{Value}\\
\hline
Platform & Ubuntu 20.04.2 LTS \\ 
\hline
Network area & $50 \times 50 m^2$ \\
\hline
Simulation time &  1200 $sec.$\\
\hline
Transmission range (sensor) & 25$m$ \\
\hline
Transmission range (mobile phone) & 50$m$ \\
\hline
Communication protocol & IEEE 802.11, 2.4 GHz WiFi\\
\hline
\end{tabular}
\end{table}

We considered three different network settings, each consisting of a varying number of patients $P$, sensor devices $SN$, and medical experts $ME$ to measure the network performance. Table \ref{Table10} describes the details of the network settings. Results presented in different graphs are averaged over several simulations.
\begin{table}
\centering
\caption{Network Settings}
\label{Table10}
\begin{tabular}{|p{1.2cm}|p{1.8cm}|p{1.2cm}|p{1.2cm}|}
\hline
{\textbf {Network scenarios}} &
 {\textbf{No. of medical experts}} & 
{\textbf{No. of patients}} & {\textbf{No. of sensors}} \\ 
\hline
1 &  1 & 1 & 1$\sim$10\\
\hline
2 &  1 & 3 & 1$\sim$10\\
\hline
3 &  3 & 3 & 10\\
\hline
\end{tabular}
\end{table}

\subsection{Impact on Throughput}
\begin{figure}
    \centering
    \subfloat[$ME$=1 and $P$=1]{\includegraphics[ width=0.50\linewidth, height=3cm]{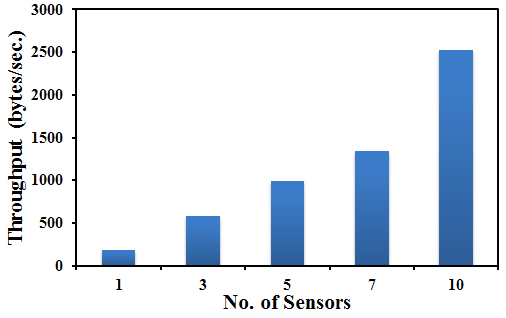}
    \label{fig:1_doc_1patient}}
    \subfloat[$ME$=1 and $P$=3]{\includegraphics[ width=0.50\linewidth, height=3cm ]{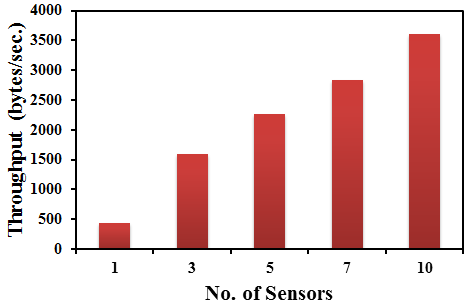}
    \label{fig:1doc_3patient}}
    \hfil
    \subfloat[$P$=3 and $SN$=10]{\includegraphics[ width=0.65\linewidth, height=3cm ]{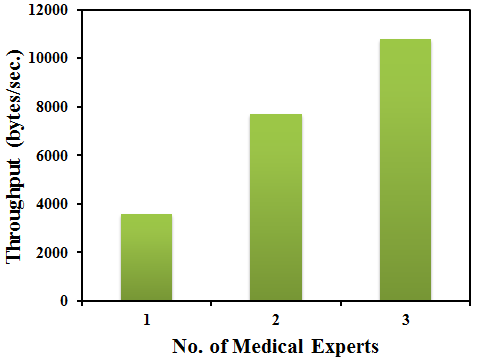}
    \label{fig:3doc_3patient}}
    \caption{Throughput for different values of $ME$, $P$ and $SN$.}
    \label{fig:throughtput}
\end{figure}

Throughput measures the number of bits transmitted per unit of time, calculated as $\frac{(N\times\lvert packet\rvert)}{T}$ where $N$ is the number of received packets, $\lvert packet \rvert$ is the size of a packet, and $T$ is the total time in seconds. Figure \ref{fig:throughtput} presents the network throughput of the proposed scheme for different network settings, where throughput increases linearly with higher $ME$, $P$, and $SN$ values. Figure \ref{fig:throughtput} (a) shows that when a single medical expert observes a patient, throughput increases linearly with a growing number of sensor nodes. Similarly, Figure \ref{fig:throughtput} (b) indicates that when a medical expert monitors three patients simultaneously, throughput also escalates with the number of patients and sensor nodes. Besides, Figure \ref{fig:throughtput} (c) shows that network throughput also rises with the increasing number of medical experts for a fixed value of $P$ and $SN$ (Here, $P$=3 and $SN$=10). 

\subsection{Impact on End-to-End Delay} 
\begin{figure}
    \centering
    \subfloat[$ME$=1 and $P$=1]{\includegraphics[ width=0.50\linewidth, height=3cm]{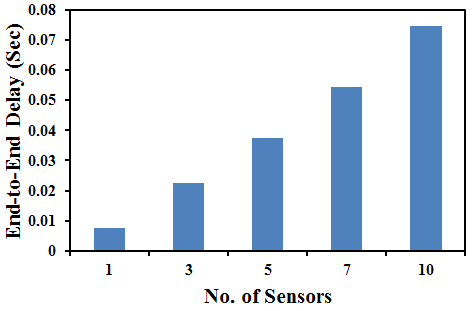}
    \label{fig:1_doc_1patient_e2e}}
    \subfloat[$ME$=1 and $P$=3]{\includegraphics[ width=0.50\linewidth, height=3cm ]{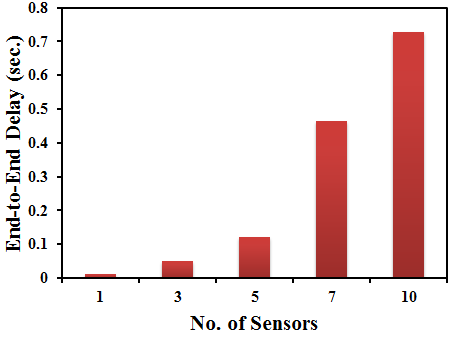}
    \label{fig:1doc_3patient_e2e}}
    \hfil
    \subfloat[$P$=3, and $SN$=10]{\includegraphics[ width=0.65\linewidth, height=3cm ]{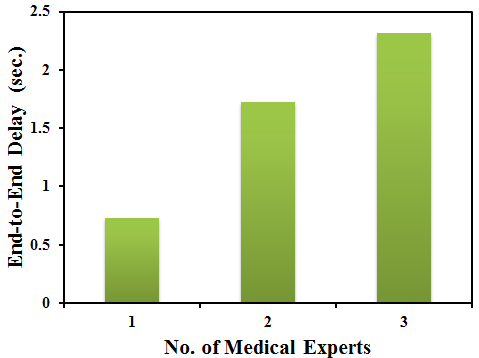}
    \label{fig:3doc_3patient_e2e}}
    \caption{End-to-end delay for different values of $ME$, $P$ and $SN$.}
    \label{fig:end-to-end delay}
\end{figure}

The end-to-end delay (EED) specifies the average time required for data packets to reach the destination from the source. It is computed as $\frac{\sum_{i=1}^{N} (T_{r_i}-T_{s_i})}{N}$, where $N$ is the total number of received packets, $T_{r_i}$ is the time when a packet $i$ reaches the destination, and $T_{s_i}$ is the time when a packet $i$ is sent from the source. Figure \ref{fig:end-to-end delay} shows the EED of the proposed scheme for various network scenarios. It is clear that EED increases linearly with growing number of $ME$, $P$, and $SN$. Figure \ref{fig:end-to-end delay} (a) indicates that EED increases with sensor nodes when $ME$ = 1 and $P$ = 1. Likewise, EED escalates with sensor nodes when $ME$ = 1 and $P$ = 3 shown in Figure \ref{fig:end-to-end delay} (b). In addition, Figure \ref{fig:end-to-end delay} (c) shows that EED grows with the number of medical experts for $P$=3 and $SN$=10. 
\section{Conclusion}\label{conclusion}
We have proposed an end-to-end authentication mechanism for WBAN that computes a secret session key between a medical expert and a specific sensor node affixed to the patient's body. Additionally, our scheme attains a comparable computation and communication cost in contrast to the related existing works. Besides, our method is resilient even if the patient's mobile phone is semi-trusted. We have performed both BAN logic analysis and informal security analysis of the proposed scheme that prove the soundness of the proposed scheme against different security attacks. Moreover, we have evaluated the influence of the proposed system on various network parameters using the NS-3 simulator and found that it obtains adequate network performance.  Hence, we believe that the proposed scheme advances the existing researches one step further and strengthen the security of WBAN.

\bibliographystyle{IEEEtran}
\bibliography{IEEEabrv,reference}
\end{document}